\documentclass[english,aps,prx,twocolumn,showpacs,superscriptaddress,floatfix]{
revtex4-1}
\usepackage[T1]{fontenc}
\usepackage[latin1]{inputenc}
\usepackage{amsmath}
\usepackage{graphicx}
\usepackage{subfig}
\usepackage{amssymb}
\usepackage{dcolumn} 
\usepackage{bm} 
\usepackage{color}
\usepackage{cleveref}

\makeatletter

\usepackage{amsfonts}
\usepackage{epsfig}
\usepackage{dsfont}

\newcommand{\be}{\begin{equation}}
\newcommand{\bea}{\begin{eqnarray}}
\newcommand{\eea}{\end{eqnarray}}
\newcommand{\ee}{\end{equation}}

\usepackage{babel}
\makeatother

\begin{document}
\title{Metastable Features of Economic Networks and Responses to Exogenous Shocks}
\author{Ali \surname {Hosseiny}}\email{al_hosseiny@sbu.ac.ir}
\author{Mohammad \surname {Bahrami}}
\affiliation{Department of Physics, Shahid Beheshti University, G.C., Evin, Tehran 19839, Iran}
\author{Antonio  \surname {Palestrini}}
\author{Mauro  \surname {Gallegati}}
\affiliation{Department of Economics, Universit\`a Politecnica delle Marche, Ancona, Italy}

\date{\today}

\begin{abstract}
It has been proved that network structure plays an important role in addressing a collective behaviour. In this paper we consider a network of firms and corporations and study its metastable features in an Ising based model. In our model, we observe that if in a recession the government imposes a demand shock to stimulate the network, metastable features shape its response. Actually we find that there is a minimum bound where demand shocks with a size below it are unable to trigger the market out from recession. We then investigate the impact of network characteristics on this minimum bound. We surprisingly observe that in a Watts-Strogatz network though the minimum bound depends on the average of the degrees, when translated into the economics language, such a bound is independent of the average degrees. This bound is about $0.44 \Delta$GDP, where $\Delta$GDP is the gap of GDP between recession and expansion. We examine our suggestions for the cases of the United States and the European Union in the recent recession, and compare them with the imposed stimulations. While stimulation in the US has been above our threshold, in the EU it has been far below our threshold. Beside providing a minimum bound for a successful stimulation, our study on the metastable features suggests that in the time of crisis there is a "golden time passage" in which the minimum bound for successful stimulation can be much lower. So, our study strongly suggests stimulations to be started within this time passage.
\end{abstract}

\maketitle

\section*{Introduction}

Addressing the causes of business cycles and its dynamics is one of the most major goals of economics. Clearly, studying network structure can reveal some unknown features in this regard. It has been shown for example that the topology of the macroeconomic networks can have serious impacts on the cascades of crises in a world-wide scheme \cite{lee}. It as well has been proved that countries with higher connectivities in inter sectoral connections have more serious avalanches in the time of crisis \cite{contreras}. In regular interaction of systems, central limit theorem states that when the number of random variables grows, the fluctuations would overlap and effectively dampen each other. It however has been proved that in the network of production, it is the topology of the network that addresses constructive or destructive fluctuation effects \cite{Acemoglu}. While some networks are robust to the random fluctuations, some other are vulnerable. A big deal of attention in the literature has been devoted to find out how networks of production are vulnerable to the propagation of crisis. In our paper, on the contrary we study the response of the network of firms to a demand shock and the recovery acts imposed by policy makers.

In the Great Recession which occurred recently, amongst economists who favoured the stimulation policies, some had the belief on the need for a very big fiscal stimulation to recover the economy. This was while economy was in a situation characterized by a zero lower bound when the near zero short-term nominal interest limited the Central Bank capacity to stimulate economic growth (see for example \cite{woodford2011simple} and \cite{lawrence2011government}). In other words they suggested that, in such an extreme situation, a small stimulation fails to help economy for fast recovery. In this paper through studying an Ising model of networks of firms, we investigate the metastability features of the network and its response to a demand shock. We actually evaluate the existence of the minimum bound for the size of a successful fiscal stimulation and its relation to the production output.  

There can be different forms of fiscal stimulation such as tax cut and Government spending. We however simplify the model and suppose that the government only make purchases from firms and corporations in a fiscal stimulation. Actually a tax cut itself will be an indirect purchase from firms and corporations.

Ising model has been proposed as a simple model to describe networks of firms (see for example Brock et al. (2001) \cite{William2001} and Durlauf et al. (2010) \cite{Durlauf}). The proposal is as follows.
 In a network of firms in an economy, each firm is connected to some other firms. Firms and corporations can buy each other's products as intermediate goods or services. Now, each firm has a maximum and minimum capacity of production. For each firm there is a minimum level where production below it results in loss rather than profit. Besides, each firm has a maximum capacity of production where producing over that is not possible. Each firm can sell its products as intermediate goods or services to its neighbours in the network of firms. Now, for a firm that its neighbours work with their maximum capacity, it is more likely that orders are high allowing working with maximum capacity. As well, if the neighbours work with minimum capacity, then it is likely that this firm works with its minimum capacity. So, firms force their neighbours to have a status similar to themselves. The situation is however stochastic and you can only talk about probability. If your neighbours do not buy your total products, you can still work with your maximum capacity and keep your extra production as inventory investment. This possibility is however limited.

If we want to model the interaction of firms in a network, we can simplify it as much as possible. As the simplest model, we can think of firms with a bi-status situation. Each firm can choose either its maximum production level or its minimum production level, what we may indicate by up and down status. In a network of firms, if your neighbours choose an up status, they force you to choose an up status and vice versa. So, a network of firms at its simplest approximation can be considered as an Ising model. For a comprehensive discussion see \cite{{Durlauf},{William2001}}. 

In a collection of firms in an interacting network we expect collective behaviours and emergent phenomena. In a Keynesian economy it is believed that in depression where unemployment is high, the agents themselves reduce consumption. As a result, the Keynesian school suggests intervention of government for economy to recover. If we consider an Ising model as a model to describe economy, then we can read the Keynesian view of a depression as the behaviour of an Ising model below critical temperature. Below the critical temperature, we have symmetry breaking. When a maximum of firms choose to work with minimum capacity, then without a shock such as government stimulation, it is unlikely that their majority would decide to change mind and work with maximum capacity. So, if we study networks of firms through an Ising model, to simulate large scale crises within the Keynesian framework, we should consider the model under temperatures below the critical temperature.  

To find if there is a minimum bound for fiscal stimulation in economy we consider metastable features of an Ising model. For an Ising network below the critical temperature, when a majority of dipoles have chosen a downward direction, in order to stimulate them to choose an upward direction, you need to impose an external magnetic field. Theoretically, we believe that even with the effect of very weak stimulating fields, the system changes its status and move from a downward direction to upward. The process however might be time consuming. If the stimulating field is very weak, then for a long period of time, spins may not change their direction to  upward. That's why we call these states, metastable states. If in our problem we are concerned about time (as we really are in economy), then the intensity of stimulating fields actually matters. In this paper we review works concerning metastable features of the Ising model before translating the results of such studies to the macroeconomy language. We then run simulations concerning both cubic and small world networks to find the desired minimum bound.

Metastable features of the Ising model have been widely studied in physics. For studies concerning kinetic Ising see \cite{Binder}. Life time of the metastable states has been worked out in \cite{stoll77}-\cite{Tomita}. Understanding kinetic behavior of the matter utilizing the droplet theory was considered in \cite{Rikvold94}. Dynamic phase transition was studied to understand some critical features of the matter in \cite{Sides99}-\cite{Fujisaka}. For a study on the response of the model to an impulse stimulating field see \cite{Acharyya} and \cite{Misra}. For a review on droplet theory and dynamics of the Ising model see \cite{{Gould}}-\cite{Chakrabartii}.

Statistical physics has taught us that the path from micro to macro is not straightforward. This is why heterogeneous agent based models are studied (see for example \cite{deligatti}-\cite{Zhang}). Ising model has been applied widely in this framework (see for example \cite{Zhou}). Critical phenomena made it clear that network structure can play an important role to address aggregate behaviour. Currently many different issues are being studied via network glasses \cite{Newman}-\cite{Espinal}.

In this paper, trying to understand metastable features of the network of firms and corporations in economy, we first translate desired parameters in economy to the parameters in the Ising model. We then review studies on the life time of the metastable states and direct our analysis towards the appropriate domain for simulation. We first check if there is a minimum bound for a successful fiscal stimulation in an Ising approximation of networks of firms in two dimensional cubic lattice. We then search for such a minimum bound in a small world network. Finally, we consider our findings and try to have an examination for economies of the US and EU in 2009 when a stimulation policy were to be imposed. Surprisingly, in spite of the serious simplification, the model suggests a reasonable bound.




\section*{Materials and Methods}
\subsection*{Metastable Features in an Ising Model}

\begin{figure}
\includegraphics[width=1.0\columnwidth]{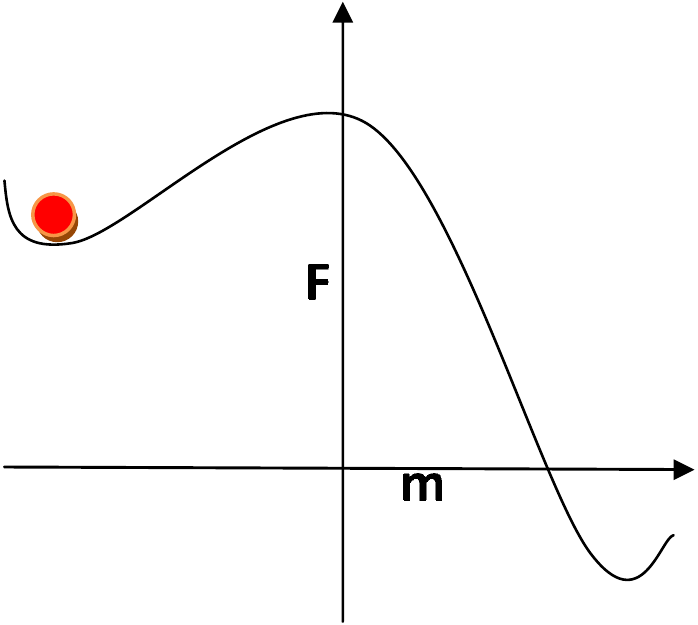}
           \caption{ Below the critical temperature, the system is in its minimum 
with $m\approx -1$. By imposing a weak upward external field, although the symmetry breaks and theoretically the system should move to its global minimum, the transition is still time consuming.} 
  \label{fig1}
 \end{figure}

Metastable features have been widely studied in the Ising model. Consider an Ising model with Hamiltonian
\begin{eqnarray}\label{j}
{\cal H}=-\beta J\Sigma_{<ij>}S_i.S_j-\beta H\Sigma_iS_i,
\end{eqnarray}

where $J$ represents the interaction of neighbors, and $H$ indicates the external field. In the absence of an external field, and below the critical temperature, the symmetry is broken and the system finds a non-zero magnetization. In other words the majority of spins choose the same direction say a downward direction. Now, if we impose an external field in the opposite direction (say upward), from the theoretical point of view the majority of the spins should flip upward putting the system into its other vacuum. Although theoretically in the presence of the external field, the free energy is no more symmetrical, and the system should fall into its global minimum, the process is still time consuming. Now let's see how it works. From a free energy point of view, after imposing an upward magnetic field, the global minimum is around $m\approx 1$, see Fig. \ref{fig1}.
 If we look at micro levels however things are different.

\begin{figure}[h]
\includegraphics[width=\columnwidth]{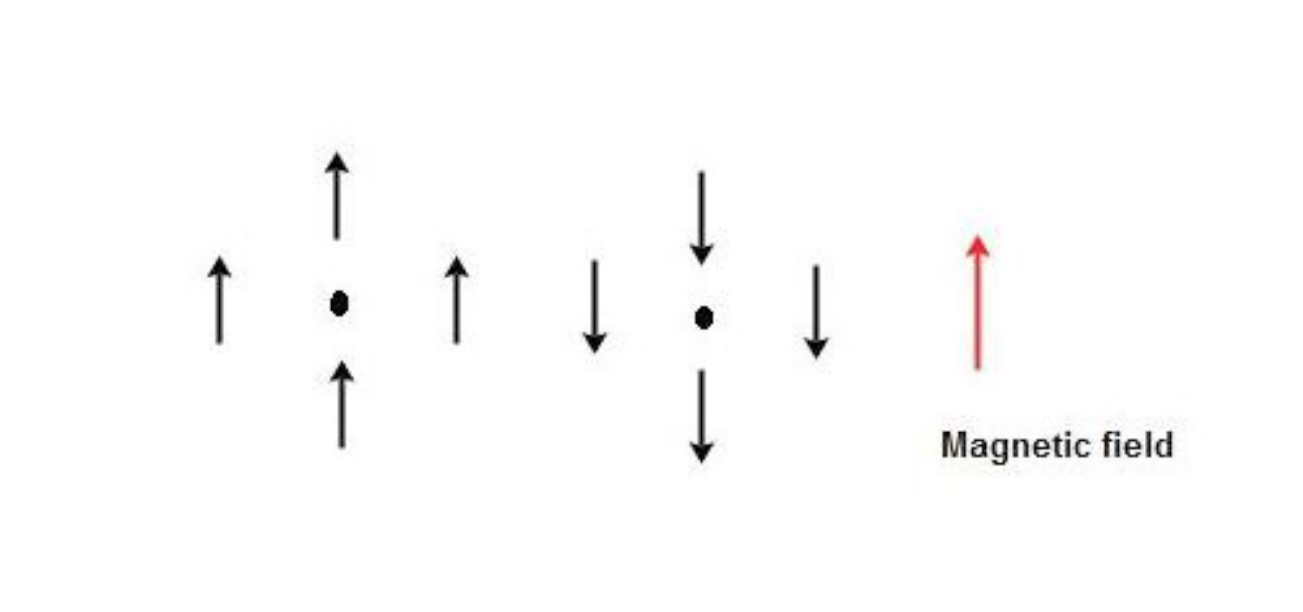}
           \caption{All spins are in downward direction as we have impose an stimulating
 upward field. Theoretically, if all spins turn upward, the level of energy will
 be in its minimum. In the meantime however if the external field is weak, 
then each spin should overcome forces of neighbors and go to a higher
 local level of energy. For strong fields things are different. If the stimulating
 field is about $8J$ for example, then the spin feels that all neighbors are
 upward and flip easily.} 
  \label{fig2}
 \end{figure}

At microlevel, for a chosen spin, in average the neighbors are downward. If the stimulating field is weaker than $4J$, choosing an upward direction for the spin would result it in having a higher level of energy, see Fig. \ref{fig2}. Due to the Boltzmann energy probability, the spin is unwilling to comply with the external field. In the language of game theory, it is similar to the prisoner dilemma problem. If all spins choose an upward direction, the total energy will be lower and every dipole is better off. For a sole spin however flipping upward is to get in a higher level of energy causing an unwanted situation. In a network of firms in recession, if all firms simultaneously decide to hire new labors and work with maximum capacity, everybody is better off. For a single firm, starting production with maximum capacity while orders are in minimum level is a risky act that most probably results in loss. From a Keynesian point of view a government stimulating policy similar to a magnetic field helps economy to escape from its unwanted minimum to a favorable one.

Let's come back to our Ising model. After imposing a magnetic field gradually more and more spins flip upward.
In our problem we are concerned with the lifetime of the metastable state. Suppose that below the critical temperature our system has chosen the downward direction with magnetization per site is close to $-1$. Now, we impose a stimulating upward magnetic field with intensity $H$. If the intensity is low, then since for each spin forces of its neighbours is bigger than the stimulating field, then the chance for it to flip is small. When time goes by, however gradually the stimulating field manages to flip more and more spins upward. When half of spins are upward, then in average for each spin the forces of neighbours cancel out outing us from a metastable trap. In this situation our stimulating field easily forces spins to take an upward direction. The period that it takes for the stimulating field to drive magnetization from its initial value to a zero value is called the metastable lifetime denoted here by $\tau$.

\begin{figure}[h]
  \includegraphics[width=\columnwidth]{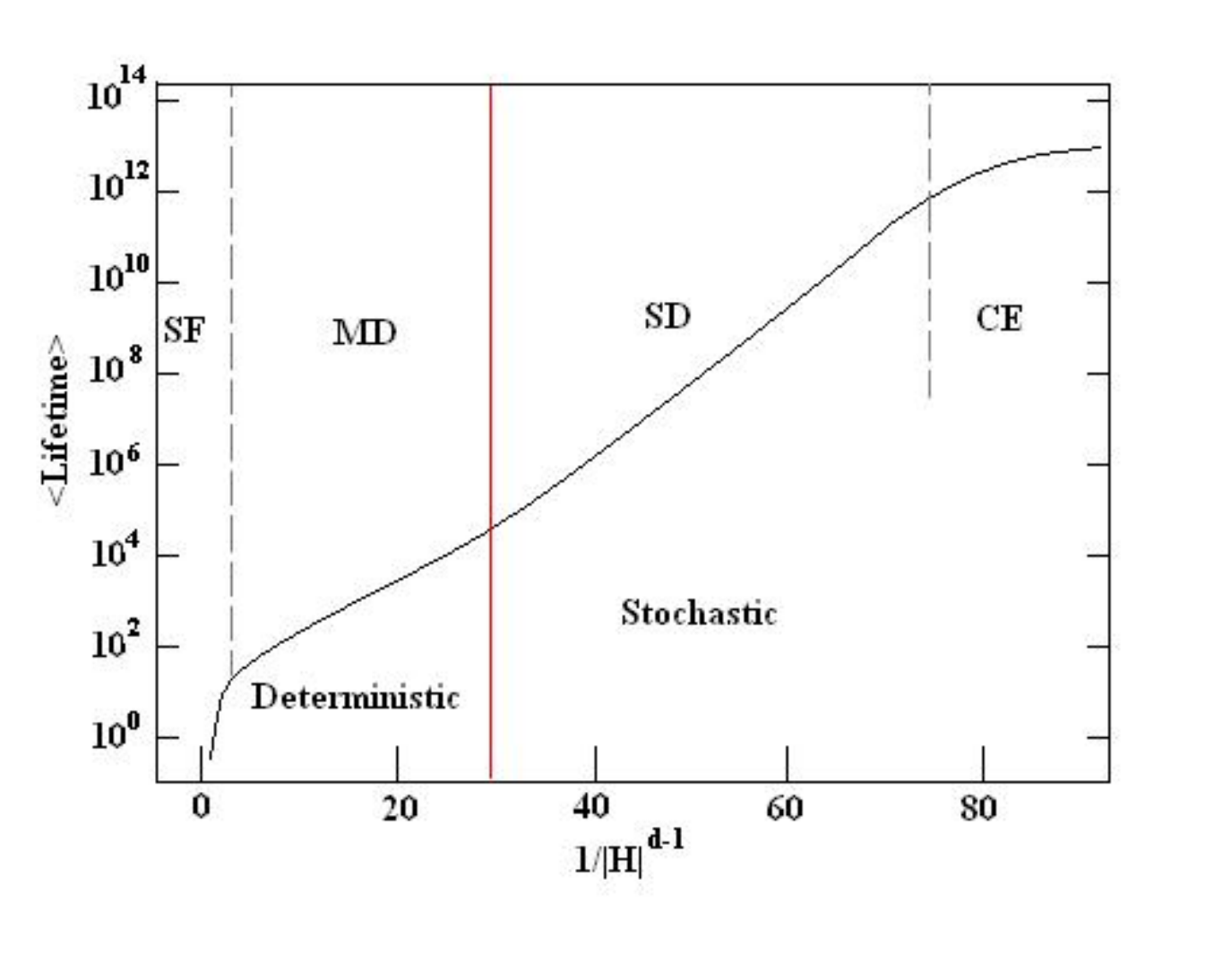}
     \caption{The schematic life time of metastable states as a response to the strength of the stimulating field from \cite{Rikvold94}. The weaker the stimulating field, the longer the life time. The first two regimes in the right are stochastic regimes in which the life time is not only long but also has serious variances. These two regimes are named as the Coexistence regime and the single droplet regime. For stronger fields we have two regimes in which transition from metastable states happens in a deterministic way with small life time variance. These two regimes are Muti-Droplet regime and Strong Field area. }
\label{fig3}
\end{figure}

 Generally, size of the system, its temperature, and the intensity of the stimulating field can influence the lifetime of the metastable states. Major studies in cited papers take the temperature around $0.8 T_c$. When we study the response of the system to the stimulating field we basically define four separate regimes. Lifetime of the metastable states has been depicted schematically in Fig. \ref{fig3} (see \cite{Rikvold94} for more details). 

Let's explain what Fig. \ref{fig3} represents. At the right side of the graph we have responses of the system to the very weak fields. When the stimulating field is too weak the lifetime of the metastable states can be dramatically high. For finite size systems in this regime, the model lives in a coexistence regime in which thermal fluctuations are even more important than the stimulating field. In this regime even if the system moves upward, through thermal fluctuation dipoles may move downward again. The regime is however a stochastic regime in which the variance of the lifetime is comparable with its mean value. Both the magnitude of the lifetime and its variance depend on the size of the system, which dramatically grows as the size grows. Such a regime is called "Coexistence" regime. 

The second regime is called single droplet regime. In this regime the stimulating field is a little bit stronger, and the system would finally go to its global minimum. The lifetime is however long and the response of the system is stochastic. We should wait for a chance for the formation of a droplet with upward spins. This droplet takes a long time to grow and capture the whole system. Although in this regime we can find an average value for the lifetime, the variance is still comparable with the average value. So, the regime is still a stochastic regime. The regime is called "Single Droplet" regime.

The third regime is where the stimulating field is strong enough to form many droplets and grow leading to escaping from the metastable trap in a predictable period. The lifetime is no more stochastic and would be  
\begin{eqnarray}\label{lifetime}
\tau\left( H \right) =c_{{1}}{{\rm e}^{{\frac {c_{{2}}}{3 \left| H
 \right| }}}}   \left| H \right| ^{-5/3},
\end{eqnarray}
in which $c_1$ and $c_2$ are constants. As we expect, the lifetime dramatically decreases as the stimulating field increases. This regime is called "Multi Droplet" regime.

The last regime is called the "Strong Field" regime. It is a regime in which the stimulating field is strong enough that each spin has a chance to flip. The rate is so high that we do not need to wait for some survivable droplets to form. 
\section*{Results}



\section*{Relation Between Macroeconomics Variables and the Ising Model}

In the previous section we reviewed the responses and the lifetime of metastable states in the Ising model. We now need to interpret these responses in the language of economy. We need to report a minimum bound and money value of a fiscal stimulation. For the beginning let's consider a cubic $D$ dimensional lattice. In $D$ dimensions each dipole has $2D$ neighbors. In a downward vacuum their force on a dipole is around $-2DJ$, see Fig. \ref{fig2}. If they all turn from a downward direction to an upward direction, their force on our dipole is equal to $2DJ$. Our stimulating field appears as $H$. In a downward vacuum each spin feels a downward force from neighbors almost equal to $-2DJ$, where if the stimulating field is about $4DJ$ each spin feels a net force equal to $2DJ$. In this case, each spin can simply suppose that all of its neighbors are upward and there is no exogenous field.

On the economy side we supposed that the transfer of goods or money between firms forces them to chose a downward or upward direction. Suppose that in an expansion when all firms and corporations work with maximum capacity, the GDP is in its maximum level which we denote as $GDP_+$. In a recession when all firms work with their minimum capacity, the GDP is in $GDP_-$ level. Let's denote the difference between this gap as $\Delta GDP=GDP_+ -GDP_-$. In a recession, when all firms work with a minimum capacity, if the government has a purchase as big as $\Delta GDP$, then it has compensated for the reduction of purchase by the neighbours. If we compare it with our Ising model, we find that a stimulation as big as $\Delta GDP$ resembles a stimulating field as big as $4DJ$.

 Stimulating bill may be designated to be spent for periods more than one year. In this case we need to write
\begin{eqnarray}\label{size}
\frac{bill}{\Delta GDP}=\frac{\tau HN}{4DJN}=\frac{\tau H}{4DJ}\;\;\Rightarrow\;\;bill=\frac{\tau H}{4DJ}\Delta GDP,
\end{eqnarray}
in which $N$ is the number of spins in the lattice or the number of nodes in a network, and $\tau$ is the time interval or Monte Carlo steps we stimulate the Ising model.

 A point that should be noted is that our transformation of the parameters for the Ising model to the macroeconomics variable is true if we suppose that the firms decide for their production level in an annual base. Actually, lots of contracts with labors are on annual basis. It however by itself does not mean that firms change their strategies in an annual base since contracts of different labors may be expired in different months. Strategies however will not change with a monthly basis. Firms can bear with fluctuations of a couple of months with their inventory investments. This discussion however is of importance in the interpretation of our findings. In the end of the paper we will recall this matter and discuss the robustness of our findings for this interpretation. For now let's suppose that firms change their strategies in an annual basis.

In metastable studies the strength of the stimulating field has been discussed. In economy however we are mainly concerned with the budget. So, not only the strength of stimulation in each year is important, but also the period in which this stimulation should be imposed is important. As a result, the term that is important for us is $\tau H$ rather than $\tau$ or $H$ themselves. If there is a minimum bound for the term $\tau H$ to move an Ising model from one of its vacuums to the other, then it means that for our network of firms there is a minimum bound for an effective stimulation. So, we seek to find the minimum of $\tau H$ in the Ising model.

As it was discussed in the previous section, to classify the lifetime of the metastable states we have four different regimes. Two first regimes where stimulating field was too weak are not of our interest. It is because the lifetime was stochastic. Neither politicians, nor policy makers are interested in a stimulation that has a stochastic lifetime. Politicians will loose their positions and policy maker will loose their reputation. The first goal of a stimulation is to trigger economy from recession to an expansion in a proper period of time. So, we leave the stochastic regimes in Fig. \ref{fig3} aside and focus on Multi-Droplet and Strong-Field regimes.

\begin{figure}
  \includegraphics[width=\columnwidth,angle=0]{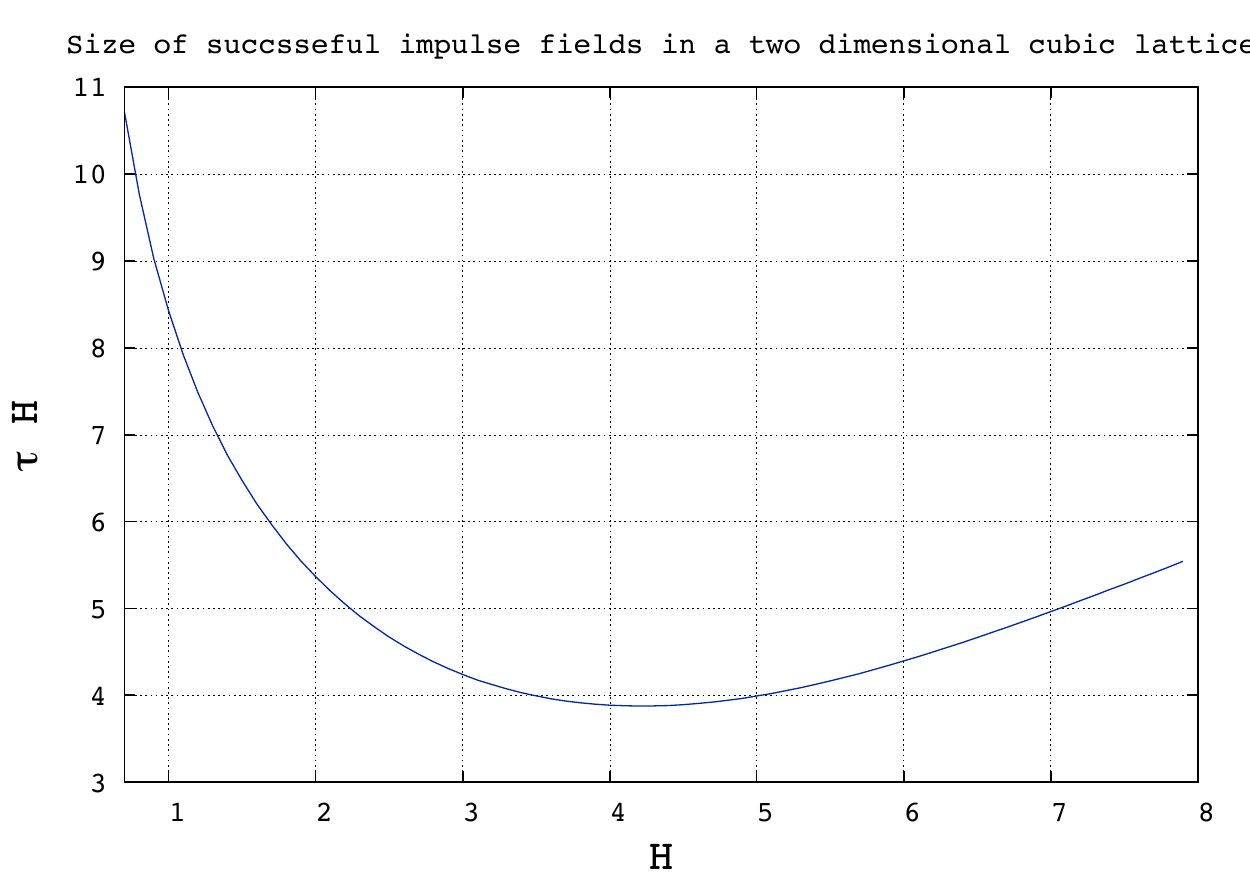}
     \caption{The result of our simulation for finding $\tau H$ for metastable states.
 H is in units of J (The coupling constant). The model is a two dimensional
 1024*1024 lattice. The current figure is for one iteration. The minimum 
of $\tau H$ is obtained (3.877$\pm$0.004)J for 100 iterations.}
\label{fig4}
\end{figure}
In the Multi-Droplet region the lifetime has been indicated in Eq. (\ref{lifetime}). Our aim is to find the minimum of $\tau H$ or $c_{{1}}{{\rm e}^{{\frac {c_{{2}}}{3 \left| H \right| }}}}   \left| H \right| ^{-2/3}$. This function is clearly a decreasing function of $H$. So, the  bigger the magnitude of the magnetic field, the smaller the $\tau H$. So, to find a minimum for $\tau H$ we are forced into the Strong-Field region. 

Response of an Ising model to a strong magnetic field has been discussed in \cite{Acharyya} and \cite{Misra}. Lifetime has been of interest in the mentioned references. Since we need to find the minimum of the term $\tau H$ rather than $\tau$ we need to run our own simulations.

 We considered a two-dimensional 1024*1024 cubic lattice at $T=0.8 T_c$ with a system living in its downward vacuum. We then imposed a magnetic field and updated each spin through the Monte Carlo method with the Glauber weight. We draw $\tau H$ as a function of $H$ in Fig. \ref{fig4}. This figure deserves attention. First of all both strong and weak fields result in higher value of $\tau H$. It is clear why weak fields result in higher rates of $\tau H$. This is because for weak fields the chance for
 every spin to flip is small. Strong fields as well results in higher values for $\tau H$ because; to flip every spin we do not need a field stronger than the effect of neighbors or $4J$. The minimum of the curve however is obtained for $H\approx2DJ$. This means that for a minimum stimulation, the government needs to impose incentives for corporations with a rate that compensates for half of the reduction of orders in the recession. In the minimum, we have $\tau H=(3.877\pm 0.004)J$. Such a field suggests a minimum for fiscal stimulation equal to  $0.48\;\Delta GDP$. For such stimulation, before we update each site once in an average, the magnetization tends to zero. In other words, the lifetime is below one Monte-Carlo step.

\begin{figure}
  \includegraphics[width=\columnwidth,angle=0]{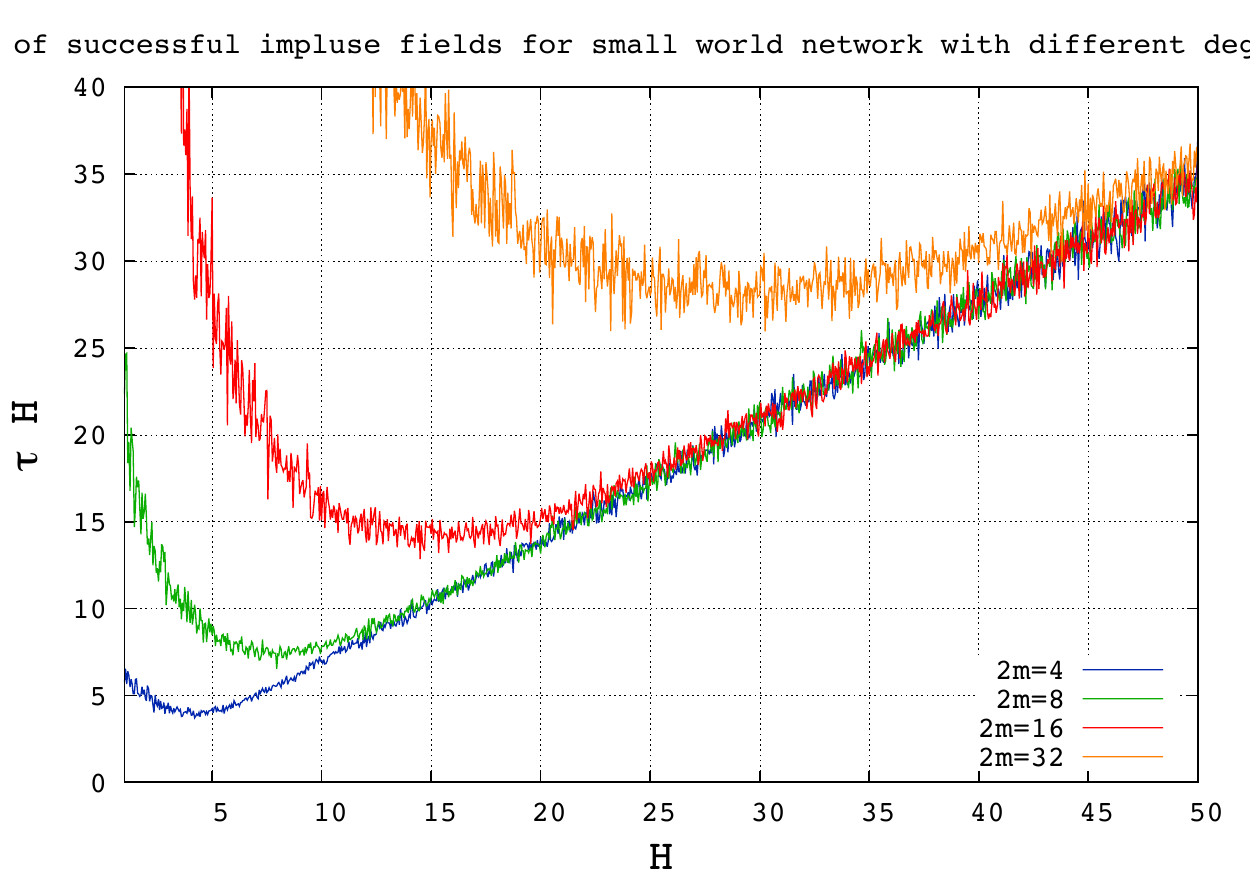}
     \caption{The result of simulation for Watts-Strogatz small world. As long as the degrees
 grows, the minimum \\
for $\tau H$ grows as well.}
\label{fig5}
\end{figure}



\section*{Studying Small World Network}

\begin{figure}
  \includegraphics[width=\columnwidth,angle=0]{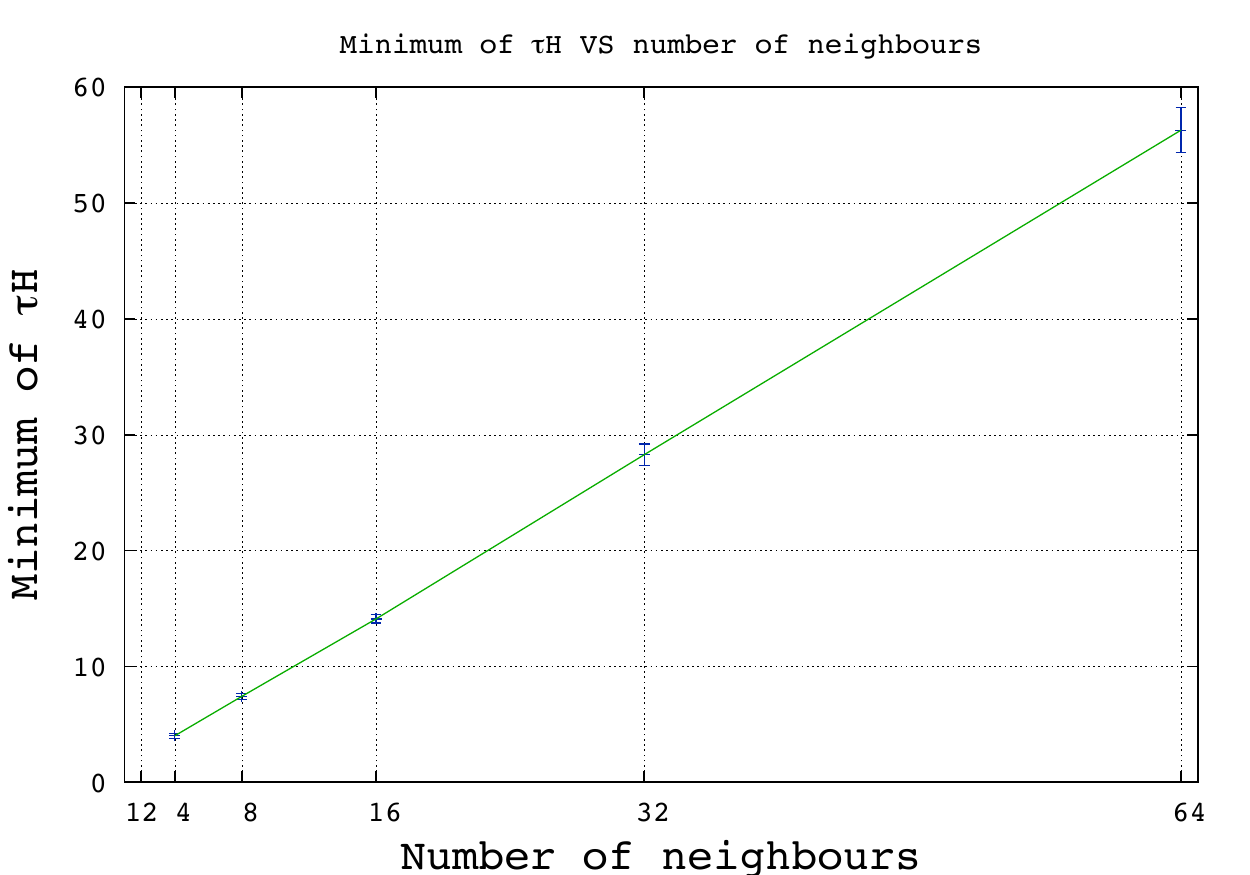}
     \caption{Minimum of $\tau H$ grows linearly with the degree of nodes. So, the ratio 
of $\tau H/2mJ$ merges to an almost constant rate.}
\label{fig6}
\end{figure}

A regular cubic lattice for the network of firms is applicable for a very simplified world.
A better approximation can be reached via going to a more realistic network. To this
aim we focused on Watts-Strogatz network. We considered networks with $2m$ neighbours and let $m$ vary from 2 to 32. The result of this simulation is depicted in Fig. \ref{fig5}. As it can be seen in the figure, as the degree of nodes grows, the minimum grows as well. Minimums of $\tau H$ as a function of $m$ has been depicted in Fig. \ref{fig6}. It can be seen from the figure that the minimum of $\tau H$ grows linearly with $m$. It is a good news for us. If we look at Eq. (\ref{size}), we notice that a linear growth of the minimum of $\tau H$ means a unique answer for a suggested bill. This means that to find a minimum for the stimulation policy we do not need to know the degree distribution of the network as long as it is a Watts-Strogatz one. Our suggestion for the minimum of the stimulating bill will be robust against the variation of the degree as long as the variation of degrees is not divergent. For large $m$, the minimum merges to $0.878\pm0.027$. So, the minimum bound for stimulation is $bill=(0.439\pm0.013)\Delta GDP$.

\begin{figure}
  \includegraphics[width=\columnwidth,angle=-0]{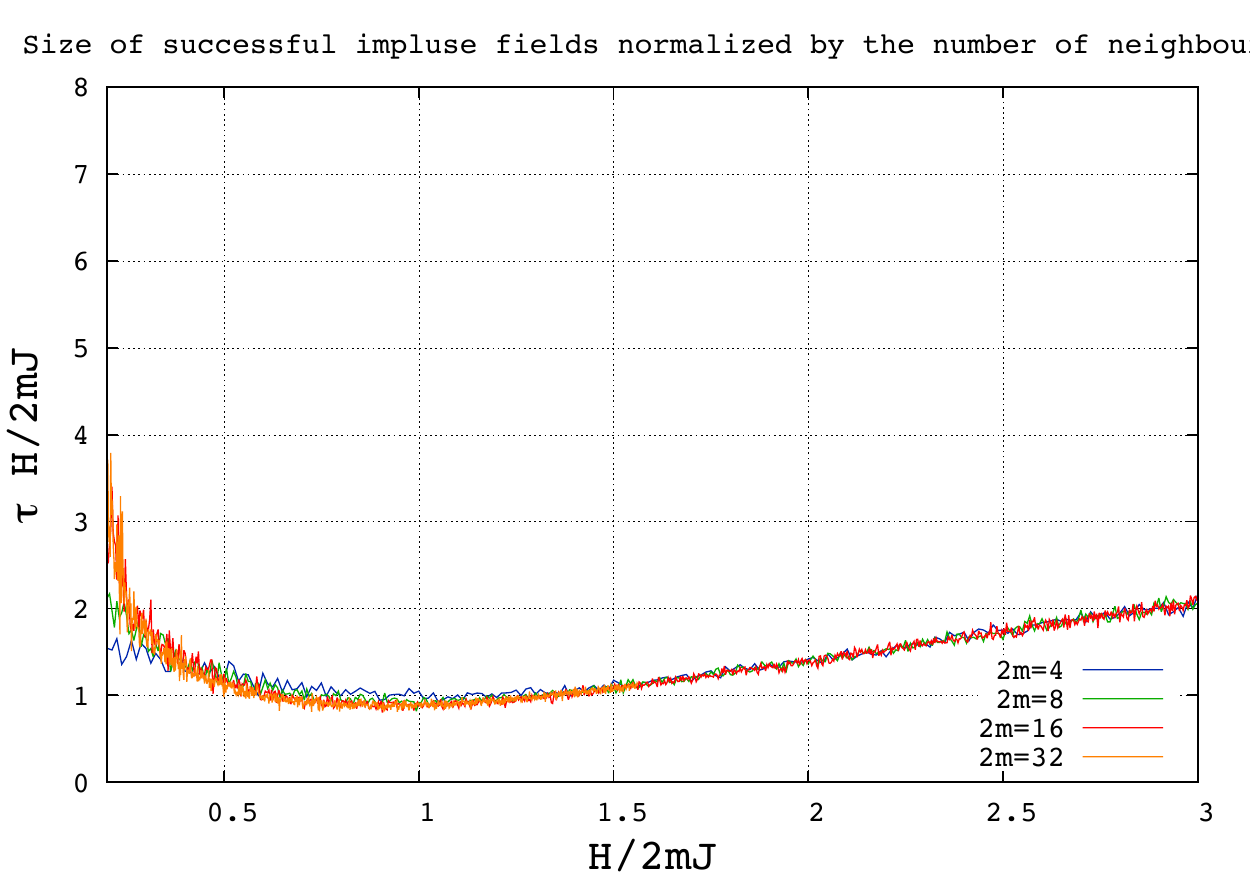}
     \caption{If in a small world we rescale the curves with the degree of nodes then all curves
 fit together. This means that all networks suggest the same minimum bound for the  
stimulating field. }
\label{fig7}
\end{figure}

Actually if in Fig. \ref{fig5} we rescale each curve by the degree of the network or $m$, we reach Fig. \ref{fig7}. As it can be seen, all curves are unified through this scaling. Let's see how it happens. In the strong field regime where we impose an intense stimulating field, after only a few Monte Carlo steps the system moves from one if metastable states to the other one. It means that with such strong fields each node does not have enough time to be influenced by far nods. Only near neighbours are important. In this case the mean field approximation works properly.

Let's suppose that in our network each node in average has $2D$ neighbours. For dynamic of magnetization we can write

\begin{eqnarray}
\begin{split}\label{meanfield}
&\frac{\partial m}{\partial t}=\frac{1}{2}[-m+\tanh[-\frac{1}{K_BT}(H-2mDJ)]
\cr&\;\;\;\;\;\;=\frac{1}{2}[-m+\tanh[-\frac{2DJ}{K_BT}(H/2DJ-m)]
\end{split}\end{eqnarray}

%

Since we supposed $T=0.8T_c$ and $T_c$ as a stationary answer of the equation proportional to $2D$, then Eq.(\ref{meanfield}) reduces to

\begin{eqnarray}\label{meanfieldd}
\frac{\partial m}{\partial t}=\frac{1}{2}[-m+\tanh[-C(H/2DJ-m)]],
\end{eqnarray}

in which C is independent of the degree of nodes. Now, if we rescale H as $h=H/2DJ$ then the equation reads as
\begin{eqnarray}
\frac{\partial m}{\partial t}=\frac{1}{2}[-m+\tanh[-C(h-m)]].
\end{eqnarray}

As it can be seen after rescaling $H$ by a factor of $2DJ$, we find a unique equation for dynamic. This is why in Fig. \ref{fig7} after rescaling the size of the stimulation by the degree of nodes we find a unique curve. So, the minimum of $\tau H$ grows with the average of the degree of nodes.

Our observation implies that in Eq. (\ref{size}) the size of the minimum successful bill is universal. In other words it is independent of the number of neighbours or degree average of nodes. We however should be careful since in our mean field argument we supposed that we could work with the average of neighbours. Our proof holds true if the standard variation of the degree of nodes is small in respect to its average. If we are working with scale free networks then we need to be more careful. We have omitted such cases at the current level.

\begin{figure}
  \includegraphics[width=\columnwidth,angle=0]{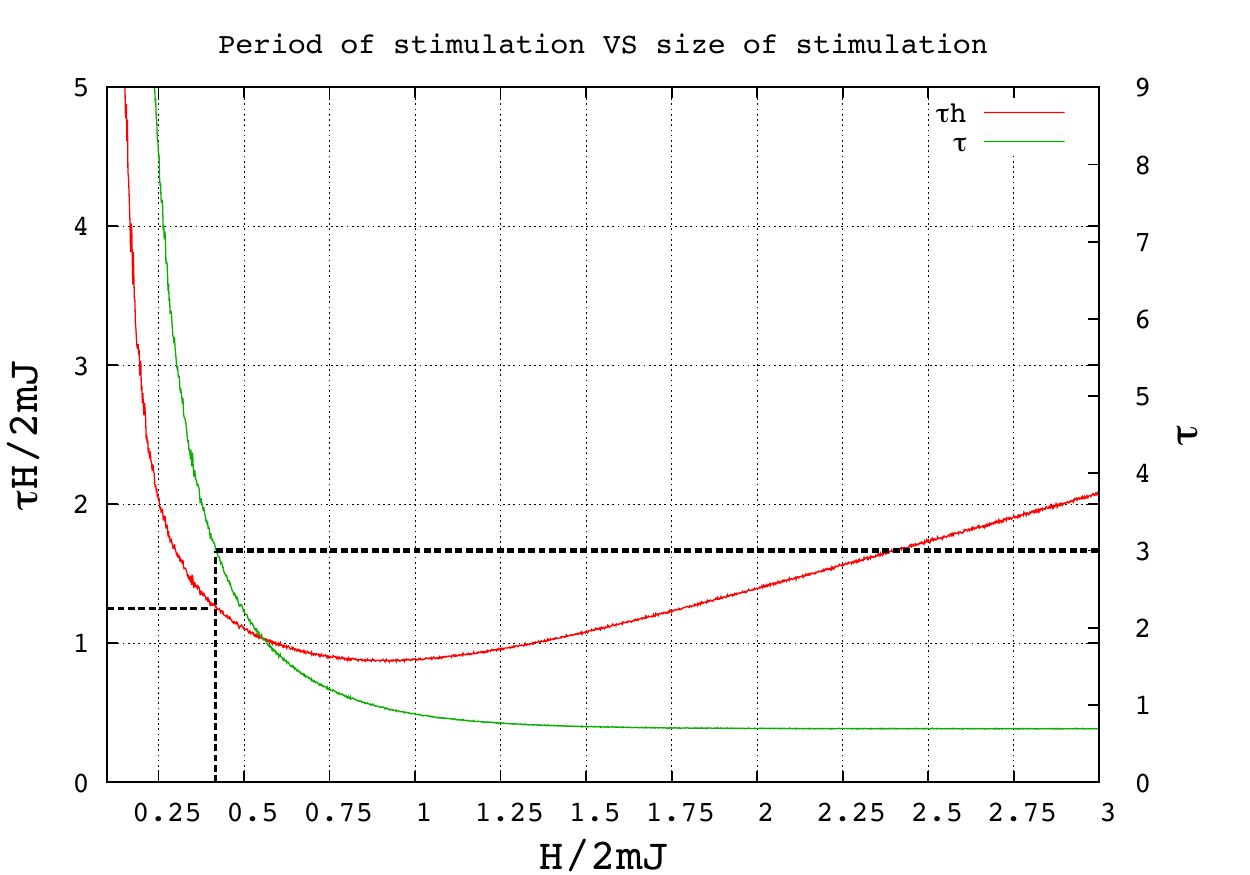}
     \caption{In this figure we have graphed both the average of $\tau H$ and $\tau$ for fifty iterations for a Watts-Strogatz network with degree of 16. Such a graph comes handy 
for the case that we have limits for budget in each year. If we aim to stimulate 
the market in periods longer than one step, then we can work with this figure in an 
Ising simulation of the network of firms. If for example instead of one Monte 
Carlo step we aim to stimulate the market in three steps, then it means that the size 
of $H$ is about $0.42$. This strength suggests the value of 1.25 for $\tau 
H$. It means that in this case, the total bill is about $0.62 \Delta 
GDP$. So for a three year stimulation, in each year we need 
a stimulation around $0.21 \Delta GDP$. For a two-year spending stimulation we
 need around 0.53$\Delta$ GDP for the whole bill and $26\% \Delta GDP$ for each year.}
 \label{fig8}
\end{figure}

{\bf Budget Constraint:}
If a government does not have enough budget to impose a big stimulation in one year, although the annual spending can be lower, the whole spending should be higher. In Fig. \ref{fig8} we see both $\tau H$ and $\tau$ in terms of the strength of stimulation. The lowest budget can be effective when stimulation is imposed for only one period and is $0.44$ of the GDP gap. If the government has a budget constraint and is willing to stimulate the economy for more than one period then this figure comes handy. The green curve shows the period that a stimulation needs to be imposed and the red curve shows the value of the total stimulus bill. If the government aims to stimulate the economy in 3 years for example, then the value of stimulation is about $0.64 \Delta GDP$, and if we have a two years plan then the minimum bound is around $0.54 \Delta GDP$.

%



\section*{Results for the Great Recession}

In this section we aim to find what is the suggestion of our model for the minimum stimulation for the US. Finding a minimum is a bit hard. It is because in the year 2009 when GDP was declining, stimulation was imposed. So, in one hand we had an avalanches of defaults and on the other hand we had stimulation being imposed. As a result extrapolating an exact value for the real gap is not possible. We however try to do our best and have at least a rough guess for the minimum bound based on our analysis. 

To extrapolate the gap of GDP for the united states we considered the time series of the real growth rate from 2000 to 2007. A linear extrapolation suggested a real growth around $2.5\%$ for 2008 and 2009. In spite of $5\%$ the growth economy of the US declined about $-0.3$ in 2008 and $-2.8\%$ in 2009. So, a rough guess for the gap is $8.1\%$. The major part of Obama's stimulus package or {\it"The American Recovery and Reinvestment Act of 2009"} was spent in 2009 and 2010. The minimum size suggested by our model for a two year period according to Fig. \ref{fig8} is around $0.54\Delta GDP$. The GDP of the US in 2007 was around 14938 B of 2009 US $\$$. So, our model suggests a minimum bound for stimulation around $bill=0.54*0.081*14938 B\$\approx650B\$$.

Obama's stimulation policy bill {\it "The American Recovery and Reinvestment Act of 2009"}
was firstly \$787 billions. Later the bill was revised to \$ 831 billions. The bill was to be 
spent between 2009 and 2019. The major part however was spent during 2009 and 2010. 
As of the second quarter of 2010 \$480 billions had been spent. As of the first quarter of
2011, around 670B\$ had been spent \cite{cea}. So, the spent bill for the first two years is a bit above our threshold. 

If we repeat such analysis for the European Union we reach to a minimum bound around $4.7\%$ of GDP. This is while {\it "The European Economic Recovery Plan"} stimulus bill was around $1.5\%$ of GDP.    

In the US, in the first quarter of 2009 according to the OECD database, unemployment was about 8.27\%. It reached to its pick in the forth quarter of 2009 equal to 9.93\%. Then it started to reduce and in the second quarter of 2011 reached to 9.07\%. So the bill was successful to reverse the growing trend of unemployment and reduce it from its peak by 1\%. The major part of the bill was spent through 2009-2010. Part of the bill however was spent in the following years. Unemployment kept its decreasing trend and the US finally managed to overcome the recession. In the EU things went in different ways. In the first quarter of 2009 unemployment was 8.27\% and reached to its local peak in the first quarter of 2010 to 9.56\%. The bill however was not able to reverse the growing trend substantially. It experienced it lowest rate in the first quarter of 2011 which was 9.39\%. It was only 0.3\% below the peak. Then unemployment grew gradually and the EU could not overcome the recession.  

For sure there has been a wide range of policies affecting economy. Extensive monetary policies held by the Central banks are of the most serious ones. To have a reliable conclusion one needs to consider a careful temporal pattern of spending in a couple of countries. Still a conclusive conclusion can not be reached except if we bring some effects such as monetary stimulations into account. So, we can not attribute the recovery of the US economy totally to the fiscal stimulation act. Nevertheless the stimulus act has been of great importance. Based on our simple model the US stimulus package was above the minimum barrier of the metastable feature of the network and the European one was far below it.  So, at least at the first level of approximation, our model has successful prediction for the two major economies of the world.


\section*{Discussion}


Considering an Ising model for the network of firms is simplifying the real world. It however sheds light on our problem and exposes existence of a minimum bound for stimulation. If the network of firms is a Watts-Strogatz network then our result is independent of the degree of nodes.  

{\bf Short range or long range interaction?} One point of strength of our method is that it lives in a regime described by the mean field. If we look at Eq.(\ref{meanfieldd}) we notice that it has deep consequences. Keynesian economics believes in animal spirit. When economy falls into a deep recession people themselves cut their consumption and hesitate risky movements. Such effect makes the crisis more disastrous. Getting out of recession will be even harder. If a stimulus bill triggers aggregate demand, as long as the unemployment decreases, more and more people start to expand their consumption. At first it seems that since in an 
Ising model we have considered only the effect of neighbors, then the global situation has been neglected. I mean one may think that since in a Keynesian economics global skim of economics affects decisions of each person or manager, then we should consider a long range interaction of forces. The point is that in a mean field regime and as it can be deduced from Eq. (\ref{meanfieldd}), the growth rate of $m$ decreases when $m$ is smaller. This means that in this regime the long range and short range interaction has the same effect. So, we do not need to modify our model even if we want to bring animal spirit into account. 

{\bf Variation over temperature:}  

The role of temperature deserves attention. There is no clear understanding about temperature in a many body system such as economy where there might be different intuitions. Yakovenko and Rosser for example have suggested the average of GDP per person as the temperature \cite{Yakovenko}. Any claim about a general definition of temperature in economy may raise some debates. In simple models however we can discuss the role of temperature. In our work we have tried to model network of firms by an Ising model. In an Ising model temperature states the level of correlation between neighbouring spins. The chance for two spins to have a different direction is given by
\begin{eqnarray}\label{boltzman}
p\propto e^{-\frac{2J}{KT}}.
\end{eqnarray}
If you have $m$ neighbours and all of them have chosen to decrease their production, in this model the chance that one keeps a high level of production is 
\begin{eqnarray}\label{boltzman}
p\propto e^{-\frac{2mJ}{KT}}.
\end{eqnarray}
If we suppose the temperature to be too small, it means that if neighbours of a firm choose to cut their production, the firm itself definitely cuts its production. It is not however the case in the real world. Such assumptions ruin the stochastic structure of decisions making of a manager as well as local fluctuations of the market. So for sure, to consider stochastic features of the market we can not decide to have a close to zero temperature. 

Now, the question is weather if we can choose a relatively high temperature to model network of firms. Actually as we have stated in the introduction, if you think within a Keynesian framework, then the temperature should be below the critical point. In this framework, it is believed that in a recession, without government intervention the life of a metastable state might be too long, and the market may live in a long recession. So, if we aim to model economics network with an Ising model in a Keynesian framework, to have metastable features, the temperature should be below the critical point. Besides, if we look at the unemployment time series of the US, we observe that economy is living in periods of high and low unemployment rates. In a work presented by Safdari et al. \cite{safdari} it has been shown that fractal dimension of unemployment time series is $0.66$. So, it means that unemployment has memory and actually has a positive correlation with its past. In hot temperatures in the Ising model, magnetization has no memory and fluctuates around zero. 
As a result, in an Ising model of economy temperatures can not be high. So, a moderate temperature below the critical point is reasonable.

\begin{figure}
  \includegraphics[width=\columnwidth,angle=0]{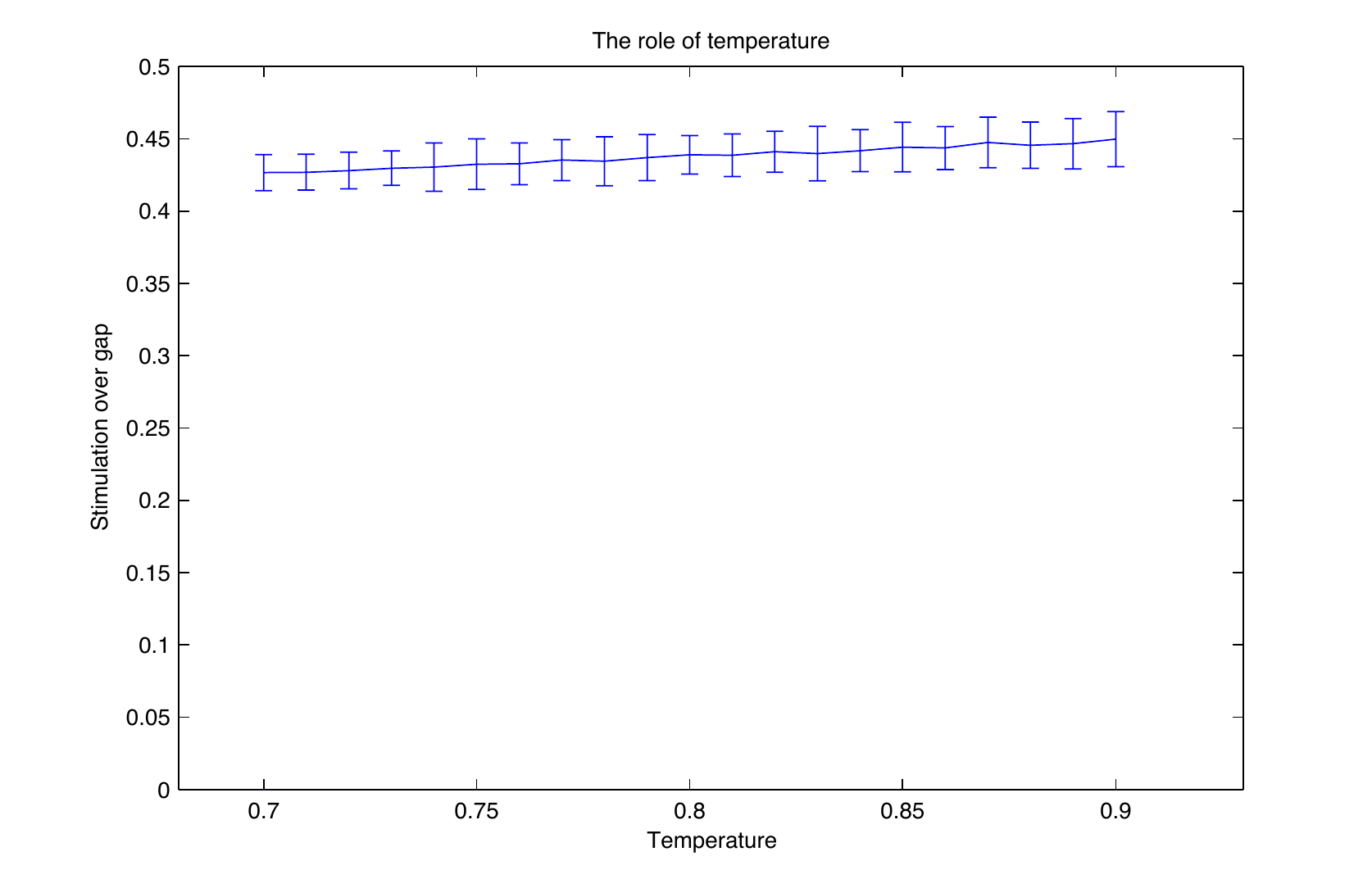}
     \caption{The minimum of $\tau H$ over different temperatures range 
from $0.7 T_c$ to $0.9 T_c$. Network is a Watts-Strogatz one
 with degree of $16$ and $P=0.01$. As it can bee seen within a reasonable window
 of temperature our result does not vary seriously. }
 \label{fig9}
\end{figure}

Now, still below the critical point, one might be concerned about the vulnerability of our results to the temperature. We ran our simulations in $T=0.8T_c$. In the introduction we have justified that if we aim to model the economy within the framework of the Keynesian economics we need to suppose that temperature is below the critical point. As far as we are concerned with qualitative discussion and metastability features, the exact temperature does not matter. In our discussion however we have translated our results into quantitative measures in economics and thereby need to check the vulnerability of our results to the temperature. In this regard we reconsidered our small world model with $K=16$ and repeated our simulation for a range of different temperatures. Results based on an ensemble of 50 iterations has been depicted in Fig. \ref{fig9} . As it can be seen in a reasonable regime of temperature our result is robust and does not seriously vary.

\subsection*{The Golden Time Passage}

Our study notified the metastable feature of the market. It is a prisoner dilemma. In recession, if a major portion of employers hire more labor and increase their production, then hired labors will buy more and everybody including managers are better off. If however only a small portion of firms do this job, they will be losers. So, in spite of the global benefits, local interests prevent the market from reaching its best situation. If we wait long, a shock such as a technological shock or demand shock may trigger the market to a global extremum, but it would be time consuming. For the Great Depression, the World 
War II rose such a demand. A fiscal expansionary policy aims to stimulate a major portion of firms to rise their production and recover the economy in a reasonable period. We found that there was a minimum bound for a successful stimulation. Our results are however vulnerable to the translating Monte Carlo steps to the real time. To reach Eq. (\ref{size}) we considered a Monte Carlo step as of one year. 

Actually if policies of managers are changed within a period of $T$, then we need to modify Eq. (\ref{size}) as
\begin{eqnarray}\label{size}
bill=\frac{\tau H}{4DJ}T\Delta GDP,
\end{eqnarray}
in which $T$ is expressed by unit of year. If managers decide to cut or rise their production in quarter base for example, the minimum bound is one fourth of our previous suggestion. The point is that in a deep recession if an employer has fired part of her employees then
 it would be unlikely that she changes her mind easily. Even if for a few months the orders 
grow, a manager may hesitate to employ new labors. If decline in production
 is due to decrease of working hour however, then it will be much easier for managers
 to rise their production. So, when we want to translate the result of our simulation 
to the language of economy, we need to be clear about the situation. In the case that
 only working hours have been decreased, the minimum bound for a successful
 stimulation can be much lower as long as the stimulation is shovel ready. In that case, 
managers can easily change their mind and rise working hours. So, an update in a Monte
 Carlo step, or actually an update for a new level of production for each agent can happen 
in a much shorter time. As a result, the minimum bound would be much lower.

\begin{figure*}
\centering
\includegraphics[width=\textwidth]{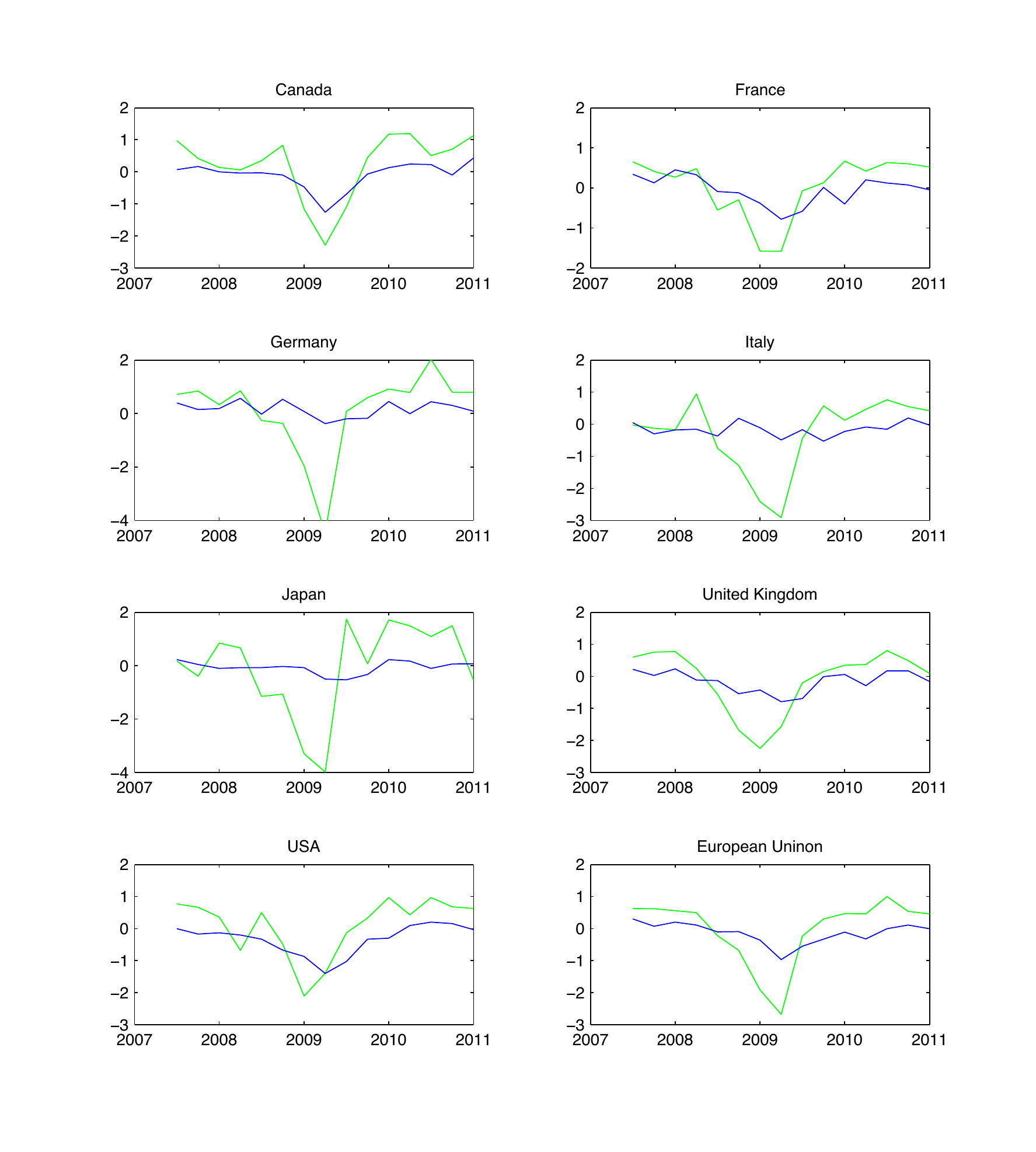}
\caption{In this figure we have depicted both changes in the GDP and unemployment. The green curve shows the GDP growth rate in quarter bases. The blue curve shows changes in unemployment in sequential quarters time -1. Unemployment and the GDP growth rate have negative correlation. Since we aimed to show the lag between these two parameters at the time of crisis we multiplied the changes in unemployment by -1. As it can be seen at the time of crisis, the GDP declines before unemployment rises. The lag between the growth rate and unemployment is the golden time passage for a stimulation. Source: OECD}
\label{fig10}
\end{figure*}

If we look over the big economies through 2008 to 2009 we observe that the decline 
in the GDP growth rate starts a couple of months before the rise in unemployment, 
see Fig \ref{fig10}. In this figure we have depicted two time series. One of the 
time series is the GDP growth rate. The other time series is the changes in unemployment 
times minus one. Changes in unemployment has negative correlation with the GDP growth 
rate under the Okun's law. In this figure we however show that in the time of crisis this
 correlation has a lag. Since graphically it was easier to show the lag of our correlated parameters, we multiplied the variation of unemployment by minus one.

As it can be seen in the figure, in almost all countries the GDP has started to decline
 a couple of months before unemployment starts to rise (or its negative form declines). Data are from OECD and
 have been rewritten in table \ref{table}. In Germany for example during the forth 
quarter of the year 2008 and the first quarter of 2009, the GDP declined $6.4\%$. In the same 
period, unemployment raised less than half a percent. During the same period while
 the GDP has declined more than $3\%$ in France, unemployment has grown less than $1.2\%$.
 In Japan as well, despite a decline around $6.7\%$, unemployment only grew around $0.5\%$. 
This observation is not surprising. In one hand lots of contracts were in annual basis and on the
 other hand it was easier for managers to decrease working hours before making decision and
 firing part of the employees.    

Now, we would notify that our observations on metastable features suggests a golden time passage for a fiscal stimulation. This time passage is the period that after a shock such as the bubble burst of 2008 firms abruptly reduce their production as a precautionary act for decline in orders. Besides, people as a precautionary act reduce their consumption. It is while firms still have not fired their employees. Since during this time passage re-rising production for managers is much easier, and a Monte Carlo step means a shorter time in the real world, a minimum bound for a successful stimulation is much lower. A rough estimation from the time series in Fig. \ref{fig10} suggests that this time passage could be around two quarters.
So, a serious lesson from studying the metastable feature is that when a shock such as 
2008 crash happens and we expect a recession as its consequence, instead of 
negotiations in congress and offices we better hurry to impose stimulation. 
In that case a much smaller stimulation may prevent economy from falling into a 
deep recession. In this time passage every single day counts. As long as time goes by
 the minimum bound for a successful stimulation rises. We should notice that at
 the time of crisis we have a percolation of the crisis between different sectors of the market.
 Such percolation itself emphasizes a fast response. Our observation however suggests that if 
imposed simulation takes place fast, even a small fiscal stimulation might prevent the recession to 
percolate. Stimulation is better to be imposed before the cut down in working hours or firing labours takes place. As a result, a relatively small but very fast stimulation in the golden time 
passage is more effective than a pretty larger one after the growth of unemployment.

\section*{Conclusions}

Keynes claimed that in the time of crisis animal spirit forces people to cut consumption. Consumption behaviour is a key debate between two mainstream schools in economics. Our work however emphasizes a totally different point. Whether final consumers cut their consumption or not, in the intermediate level, network of firms and corporations have their own problems. They can not look at aggregate indexes and cut or increase their production. If unemployment decreases in Florida and some other east states, service providers can not hire new labors in California. Each firm or corporation should look at its neighbors in the network and decide whether to hire new labors or not. This results in the occurrence of metastable behavior in the network of production. There is friction for finding new connections. When you just need to play with local connections in your network, metastable features emerge as an inevitable result. 

Keynesian economists favor fiscal and monetary stimulation for recovery. Our analysis of metastable features of the network however suggests that if policy makers decide to impose a demand shock, then there is a minimum bound for such stimulation to be successful. While we worked with a simple model, the result should have a sense of reality though rough at the current stage. We provided an early guess for this minimum bound based on the depth of the crisis and the production gap. Our suggestion is valid as long as the network of firms is Watts-Strogatz. Network of firms have shown scaling degree distributions \cite{Aoyama}, therefore a future work on a more realistic network should provide a better understanding of the response of the network of firms and corporations.  

In the United States, the stimulus package was around our minimum bound where the US managed to overcome the recession. In the European Union, the stimulus package was far below our minimum bound where the European Union has not still overcome the crisis. Although having two successful predictions is not a noticeable confirmation of an analysis, it however is satisfying at the first level of examination. A thorough examination needs a careful analysis of the temporal pattern of spending in different countries which can be of a future work.

In the current work we considered a simple model. In the future we should develop the
 model to have better resolution over the matter. Currently evaluating the response of the
 market to stimulation is being studied from different points of views. In an interesting work for example Gallegati, Landini, and Stiglitz have shown that inequality has an adverse effect on the multiplying factor of the market \cite{gallegati2016}. So, it is expected that in the future, through studies of the collective behavior from different points of views, we have better resolution over the response of the network of agents to the government role.

In this work, for simplicity we have supposed that nearly all firms work with maximum capacity in expansion and a minimum capacity in recession. This is not the case in the real world. In a recession still a portion of firms work with maximum capacity. As well, firms are not bound to have a binary choice. They can decrease their production with different levels. So, a Potts or XY model may describe the case more realistic. It can be of a future work. As well a spin glass explanation can let different vacuums for the model. All of these extensions however bring more complications to the model as they improve heterogeneity. 

One of the most important findings of our analysis is suggestion for the golden time
 passage. Although everybody has an intuition that at the time of crisis an early response 
by the government will be important because it may prevent the propagation of the crisis,
 our analysis reveals another important point. 
Our analysis reveals that the pattern of making decision by individuals has a 
serious impact on the minimum bound of the successful stimulation. In the network language, the response is normalized by Monte-Carlo time intervals.
As a result, if we are sure that the cascade is not going to spread any more, still
there would be a serious difference between a situation where only 
working hours have been declined in the firms
and a situation where labors have been fired.
 In the first case a minimum bound for a successful stimulation is much lower. It in some senses resembles hysteresis. Before the system relaxes to the new level of supply and demand thresholds, reversing the effect has a much less price.

\begin{table*}
\begin{center}\begin{tabular}{|l|c|c|c|c|c|c|c|c|c|r|}
\hline
 && 2008I&2008II&2008III&2008IV&2009I&2009II&2009III&2009IV\\ \hline
\;Canada&Growth Rate&0.06&0.35&0.83&-1.16&-2.28&-1.10&0.45&1.18\\ \cline{2-10}
& Unemployment & 5.97 &6.00&6.10&6.57&7.83&8.53&8.60&8.47\\ \hline

\;\;France&Growth Rate&0.48&-0.55&-0.30&-1.58&-1.58&-0.07&0.13&0.67\\ \cline{2-10}
& Unemployment & 6.84 &6.93&7.05&7.43&8.21&8.79&8.78&9.18\\ \hline

Germany&Growth Rate&0.85&-0.26&-0.37&-1.96&-4.45&0.07&0.59&0.91\\ \cline{2-10}
& Unemployment & 7.92 &7.95&7.42&7.35&7.73&7.93&8.11&7.66\\ \hline

\;\;\;\;Italy&Growth Rate&0.94&-0.75&-1.29&-2.42&-2.91&-0.45&0.57&0.12\\ \cline{2-10}
& Unemployment & 6.51 &6.88&6.70&6.81&7.30&7.47&8.00&8.23\\ \hline

\;\;Japan&Growth Rate&0.66&-1.15&-1.07&-3.30&-3.39&1.74&0.07&1.71\\ \cline{2-10}
& Unemployment & 3.90 &3.97&4.00&4.07&4.57&5.10&5.43&5.20\\ \hline

\;\;\;\;UK&Growth Rate&0.06&0.35&0.83&-1.16&-2.28&-1.10&0.45&1.18\\ \cline{2-10}
& Unemployment & 5.14 &5.27&5.81&6.24&7.03&7.72&7.73&7.67\\ \hline

\;\;\;\;US&Growth Rate&0.25&-0.56&-1.68&-2.25&-1.57&-.20&0.15&0.35\\ \cline{2-10}
& Unemployment & 5.00&5.33&6.00&6.87&8.27&9.30&9.63&9.93\\ \hline

\;\;\;\;EU&Growth Rate&0.50&-0.22&-0.68&-1.92&-2.68&-0.23&0.30&0.47\\ \cline{2-10}
& Unemployment & 6.75 &6.85&6.94&7.30&8.27&8.82&9.15&9.26\\  \hline
\end{tabular}\end{center}\caption{\bf{GDP growth rate and unemployment rate of G7 and the E.U. Source: OECD}}
\label{table}
\end{table*}


\begin{thebibliography}{10}





\bibitem{lee}
 Lee, K-M, Yang, J-S., Kim, G., Lee, J., Goh, K-II., Kim, I-M.  "Impact of the Topology of Global Macroeconomic Network on the Spreading of Economic Crises.",  PLoS ONE 6, e18443, (2011)




\bibitem{contreras}
Contreras, M.G. A.,  and Fagiolo, G., "Propagation of Economic Shocks in Input-Output Networks: A Cross-Country Analysis." Phys. Rev. E 90, (2014), 062812.



\bibitem{Acemoglu}
 Acemoglu, D,  Carvalho, V., Ozdaglar, A. and Tahbaz-
Salehi, A.  "The network origins of aggregate fluctuations.", Econometrica 80, (2012), 1977-2016



\bibitem{William2001}
  Brock, W. A., and Durlauf, S.N. "Discrete choice with social interactions." The Review of Economic Studies 68, no. 2 (2001): 235-260.



\bibitem{woodford2011simple}
Woodford, M. "Simple Analytics of the Government Expenditure Multiplier".
American Economic Journal: Macroeconomics 3.1 (2011): 1-35.
   

\bibitem{lawrence2011government}
Lawrence, C., Eichenbaum, M. and Rebelo, S. "When Is 
the Government Spending Multiplier Large". Journal of Political Economy 
119.1 (2011): 78-121.




\bibitem{William2001}
  Brock, W.A., and Durlauf, S.N. "Discrete choice with social interactions." The Review of Economic Studies 68, no. 2 (2001): 235-260.



\bibitem{Durlauf}
  Durlauf, S.N., and Ioannides, Y.M. "Social interactions." Annu. Rev. Econ. 2, no. 1 (2010): 451-478.


\bibitem{Binder}
K. Binder, in "Phase Transitions and Critical Phenomena", edited
by C. Domb and M. S. Green (Academic, New York, 1976),
Vol. 58


\bibitem{stoll77}
Stoll, E. P., and Schneider, T., "Size dependence of the lifetime of metastable states in the kinetic one-spin-flip Ising model." Physica B+ C 86 (1977): 1419-1420.

   
\bibitem{ray1990}
Ray, T. S., and Wang, J-S. "Metastability and nucleation in Ising models with Swendsen-Wang dynamics." Physica A: Statistical Mechanics and its Applications 167.3 (1990): 580-588.



\bibitem{Stauffer}
Stauffer, D. "Ising droplets, nucleation, and stretched exponential relaxation." International Journal of Modern Physics C 3.05 (1992): 1059-1070.




\bibitem{Langer}
Langer, J. S. "Statistical theory of the decay of metastable states." Annals of Physics 54.2 (1969): 258-275.



\bibitem{Tomita}
Tomita, H., and  Miyashita, S."Statistical properties of the relaxation processes of metastable states in the kinetic Ising model." Physical Review B 46.14 (1992): 8886.




\bibitem{Rikvold94}
Rikvold, P.A. Tomita, H., Miyashita, S., Sides, S.W. "Metastable lifetimes in a kinetic Ising model: dependence on field and system size." Physical Review E 49.6 (1994): 5080.



\bibitem{Sides99}
Sides, S. W., Rikvold, P.A., and Novotny, M.A. "Kinetic Ising model in an oscillating field: Avrami theory for the hysteretic response and finite-size scaling for the dynamic phase transition." Physical Review E 59.3 (1999): 2710.


\bibitem{Korniss00}
Korniss, G., White, C.J., Rikvold, P.A., Novotny, M.A., "Dynamic phase transition, universality, and finite-size scaling in the two-dimensional kinetic Ising model in an oscillating field." Physical Review E 63.1 (2000): 016120.




\bibitem{Fujisaka}
Fujisaka, H., Tutu, H., and Rikvold, P.A. "Dynamic phase transition in a time-dependent Ginzburg-Landau model in an oscillating field." Physical Review E 63.3 (2001): 036109.

   





\bibitem{Acharyya}
Acharyya, M., and Chakrabarti, B.K. "Response of Ising systems to oscillating and pulsed fields: Hysteresis, ac, and pulse susceptibility." Physical Review B 52.9 (1995): 6550.





   
\bibitem{Misra}
Arkajyoti, M., and Chakrabarti, B.K.,"Spin-reversal transition in Ising model under pulsed field." Physica A: Statistical Mechanics and its Applications 246.3 (1997): 510-518.




\bibitem{Gould}
Gould H. and J. Tobochnik, {it Statistical and Thermal Physics with Computer Applications},  Princeton University Press, Princeton, NJ, (2010)


\bibitem{Chakrabartii}
Chakrabarti, B.K. and M. Acharyya, "Dynamic transitions and hysteresis", Review of Modern Physics, 71, 847 (1999)


   
\bibitem{deligatti}
 Gatti, D.D., Di Guilmi, C., Gaffeo, E., Giulioni, G., Gallegati, M., Palestrini, A.
"A new approach to business fluctuations: heterogeneous interacting agents, scaling laws and financial fragility", Journal of Economic behavior \& organization 56 (4), 489-512 (2005)

   
\bibitem{Battiston}
Battiston, S., Gatti, D.D., Gallegati, M., Greenwald, B., Stiglitz, J.E. "Liaisons dangereuses: Increasing connectivity, risk sharing, and systemic risk", Journal of Economic Dynamics and Control 36 (8), 1121-1141 (2012)

   
\bibitem{Catullo}
Catullo, E., Gallegati, M., Palestrini, A. "Towards a credit network based early warning indicator for crises", Journal of Economic Dynamics and Control 50, (2015), 78-97	
   
\bibitem{Schweitzer}
Schweitzer, F., Fagiolo, G., Sornette, D., Vega-Redondo, F., Vespignani, A. and White, D. R. "Economic networks: The new challenges.",
Science, 325, 422 (2009)

   
\bibitem{Zhang}
Zhang, X., Feng, L., Zhu, R. and Stanley, H.E. "Applying temporal network analysis to the venture capital market." The European Physical Journal B 88, 1-7 (2015)


  


\bibitem{Zhou}
Zhou, WX, Sornette D.,  "Self-organizing Ising model of financial markets",
The European Physical Journal B 55, 175-181 (2007)
\bibitem{Newman}
Newman, M, Barabási, A.L., Watts, D.J. {\it Classes of small-world networks},
Princeton University Press (2006)



\bibitem{Amaral}
Amaral, L.A.N., Scala, A., Barthelemy, M., Stanley, H.E."Classes of small-world networks", Proceedings of the national academy of sciences 97, 11149-11152 (2000)



   
\bibitem{Matteo}
 Di Matteo, T., Aste,T., Mantegna, R.N. "An interest rates cluster analysis."Physica A: Statistical Mechanics and its Applications 339, 181-188 

   
\bibitem{Onnela}
Onnela, J-P, Kaski, K. and Kert{\'e}sz, J. "Clustering and information in correlation based financial networks", The European Physical Journal B-Condensed Matter and Complex Systems, 38, 353-362 (2004)
\bibitem{Nekovee}
Nekovee, M., Moreno, Y., Bianconi, G. and Marsili, M. "Theory of rumour spreading in complex social networks", Physica A: Statistical Mechanics and its Applications, 374, 457-470, (2007)

   
\bibitem{Gurtner}
Gurtner, G., Vitali, S., Cipolla, M., Lillo, F., Mantegna, R.N., Miccich{\`e}, S. and Pozzi, S. "Multi-scale analysis of the European airspace using network community detection", PloS one, e94414 (2014)

   
\bibitem{pastor}
Pastor-Satorras, R. and Vespignani, A. "Epidemic spreading in scale-free networks" Physical review letters, 86, 3200 (2001)


\bibitem{Baxter}
Baxter, G.J., Dorogovtsev, S.N., Goltsev, A.V. and Mendes, J.F.F. "Avalanches in multiplex and interdependent networks" Physical review letters 109, 248701, (2012)	

   
\bibitem{Espinal}
 Jes{\'u}s, E.-E. and L. Hern{\'a}n. "Analysis of M{\'e}xico's Narco-War Network", PloS one, 10, e0126503, (2015)

   
   









\bibitem{cea}
Council of Economic Advisers, "The Economic Impact of the American Recovery and Reinvestment Act of 2009", July 1, 2011, Table I $\&$ II



\bibitem{Yakovenko}
Yakovenko, V. and  J. Rosser, "Colloquium: Statistical mechanics of money, wealth, and income" Reviews of Modern Physics, 81, 1703 (2009)




\bibitem{safdari}
Safdari, H., A. Hosseiny, S.V. Farahani, G.R. Jafari, "A picture for the coupling of unemployment and inflation", Physica A: Statistical Mechanics and its Applications 444, 744-750 (2016)




\bibitem{Aoyama}
Aoyama, H., Y. Fujiwara, Y. Ikeda, H. Iyetomi, and W. Souma, "Econophysics and
Companies: Statistical Life and Death in Complex Business Networks", (Cambridge, Cambridge University Press 2010) 



\bibitem{gallegati2016}
Gallegati, M.,  S. Landini, J. Stiglitz, "The Inequality Multiplier", To be appeared 







\end{thebibliography}
\end{document}